\documentclass{article}
\usepackage{dcase2022,amsmath,graphicx,url,times,booktabs, tabularx, color, bbding, amssymb, float}
\usepackage[ruled,linesnumbered]{algorithm2e}

\title{A HYBRID SYSTEM OF SOUND EVENT DETECTION TRANSFORMER AND FRAME-WISE MODEL FOR DCASE 2022 TASK 4}

\name{Yiming Li$^{1,2}$,
      Zhifang Guo$^{1,2}$,
      Zhirong Ye$^{1,2}$, 
      Xiangdong Wang$^{1,\dagger}$,
      Hong Liu$^{1}$, 
      Yueliang Qian$^{1}$, 
      }
\secondlinename{	  
      Rui Tao$^{3}$, 
      Long Yan$^{3}$,
      Kazushige Ouchi$^{3}$
      }
\address{$^1$ Beijing Key Laboratory of Mobile Computing and Pervasive Device,\\
Institute of Computing Technology, Chinese Academy of Sciences, Beijing, China, \\
eamon.y.li@gmail.com, \{guozhifang21s, yezhirong19s, xdwang, hliu, ylqian\}@ict.ac.cn\\          
        $^2$ University of Chinese Academy of Sciences, Beijing, China\\ 
        $^3$ Toshiba China R\&D Center, Beijing, China,\\ 
        \{taorui, yanlong\}@toshiba.com.cn, kazushige.ouchi@toshiba.co.jp\\
 }
 
\begin{document}

\ninept
\maketitle

\begin{sloppy}

\begin{abstract}
In this paper, we describe in detail our system for DCASE 2022 Task4. The system combines two considerably different models: an end-to-end Sound Event Detection Transformer (SEDT) and a frame-wise model, Metric Learning and Focal Loss CNN (MLFL-CNN). The former is an event-wise model which learns event-level representations and predicts sound event categories and boundaries directly, while the latter is based on the widely-adopted frame-classification scheme, under which each frame is classified into event categories and event boundaries are obtained by post-processing such as thresholding and smoothing. For SEDT, self-supervised pre-training using unlabeled data is applied, and semi-supervised learning is adopted by using an online teacher, which is updated from the student model using the Exponential Moving Average (EMA) strategy and generates reliable pseudo labels for weakly-labeled and unlabeled data. For the frame-wise model, the ICT-TOSHIBA system of DCASE 2021 Task 4 is used. Experimental results show that the hybrid system considerably outperforms either individual model, and achieves psds1 of 0.420 and psds2 of 0.783 on the validation set without external data. The code is available at \emph{\textcolor[RGB]{255,0,255}{https://github.com/965694547/Hybrid-system-of-frame-wise-model-and-SEDT}}.
\end{abstract}


\section{Introduction}
\label{sec:intro}

Sound Event Detection (SED) aims at identifying the category of foreground sound events as well as their corresponding onset and offset timestamps. Task4 of the DCASE challenge has been focusing on weakly supervised SED for several years. The DCASE 2022 Task4 \cite{dcase2022task4} is a follow up of last year's challenge \cite{dcase2021task4}. This year, in addition to exploring a heterogeneous development dataset containing unlabeled data, synthetic data and weakly labeled data, participants are allowed to incorporate external dataset or pre-trained embeddings. As last year, the SED system will be evaluated by Polyphonic Sound Detection Score (PSDS) \cite{bilen2020framework} under two different real-life settings.

For weakly supervised SED, most existing works follow the Multiple Instance Learning (MIL) framework, and formulate SED as a seq2seq classification task. They usually design  Convolutional Neural Networks (CNNs) or Convolutional Recurrent Neural Networks (CRNNs) to obtain frame-level classification probability and then apply pooling mechanism to aggregate frame-level predictions to event-level results. However, such methods do not take sound events as a whole, which may ignore some global information, such as the correlation between frames or event duration. Recently, an event-wise model, namely SEDT, is proposed to handle such problems \cite{ye2021sound}. It models SED as a set prediction problem, which directly maps audio spectrogram to a set of candidate events, thus freeing SED models from trivial post-processing, namely frame-level thresholding or median filtering. Empirical study has shown that SEDT can achieve competitive performance compared with its frame-wise counterparts \cite{ye2021sound}. Moreover, we find that the two models can supplement each other, as they solve the SED task in different ways. Therefore, combining them together may be an intuitive approach to reach promising SED performance. 

\vspace{-0.4mm}
In this paper, we describe our system participating in DCASE 2022 Task 4. It is a combination of SEDT and frame-wise CNN model. For SEDT, specially-designed training formulas, including supervised learning, self-supervised learning and semi-supervised learning, are studied to help it learn from the heterogeneous development dataset. For frame-wise CNN model, metric learning is applied to narrow the domain gap between real and synthetic data, mean-teacher framework is implemented to provide supervision for unlabeled data and a tag-conditioned CNN model is used to generate final predictions based on audio tags. After obtaining each well-trained model, we explore the fusion strategy and post-processing methods of the ensemble model. By using the methods above, the hybrid system achieves competitive results on the validation dataset.

\vspace{-2.8mm}
\section{Semi-supervised SEDT}
\label{sec:ss-sedt}
\vspace{-1.2mm}
\subsection{Sound Event Detection Transformer}
\begin{figure*}[htp]
  \setlength{\belowcaptionskip}{-0.5cm} 
  \centering
  \centerline{\includegraphics[width=0.875\linewidth]{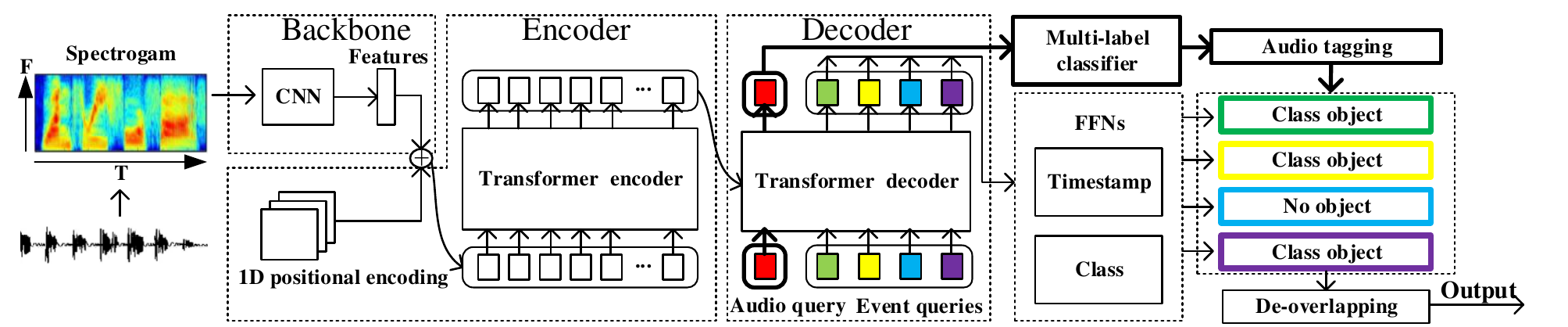}}
  \caption{Overview of Sound Event Detection Transformer}
  \label{fig:sedt_overview}
\end{figure*}
\vspace{-1.2mm}
An overview of SEDT is shown in Fig.~\ref{fig:sedt_overview}. It represents each sound event as $y_i = (c_i, b_i)$ , where $c_i$ is the event category and $b_i = (m_i, l_i)$ denotes the event temporal boundary containing normalized event center $m_i$ and duration $l_i$, and directly seeks a mapping between input features and ground-truth events. Given the input spectrogram, the backbone CNN is adopted to extract its feature map, which is then added with one-dimensional positional encoding and fed into transformer encoder for further feature processing. The transformer decoder takes $N + 1$ learnable embeddings ($N$ event queries and 1 audio query) as input event query, where each of them gathers information of a potential event from the encoder output feature via encoder-decoder cross-attention mechanism to generate event-level representations, while audio query gathers the whole audio information to generate clip-level representations. Finally, Feed Forward Networks (FFNs) are utilized to transform the event-level representations and clip-level representations from the decoder into event detection and audio tagging results, which are then fused together to get the candidate detection results. De-overlapping is implemented on overlapped candidate events of the same category. Specifically, it only reserves the events with the highest class probability. More details can be found in \cite{ye2021sound}.

\subsection{Supervised learning for SEDT}
\label{ssec:sedt training}
SEDT incorporates event-level loss and clip-level loss to optimize its event detection and audio tagging performance. For strongly-labeled data, both loss terms will be involved during the SEDT model training, while for weakly-labeled data, the event-level loss will be excluded since the strong annotations are not available.\\
\textbf{Event-level loss.} SEDT adopts a label assignment scheme before computing event-level loss: it tries to find a matching $\hat {\sigma_i}$ between each event prediction $ \hat {y_i}$ and its corresponding ground-truth annotation $y_i$ through Hungarian algorithm, which is efficient for above bipartite graph matching problem. 
To equip SEDT with sound event classification and localization ability, the loss for SEDT supervised training is formulated as the weighted linear combination of localization loss $\mathcal{L}_{\mathrm{loc}}$  and classification loss $\mathcal{L}_{\mathrm{cls}}$. For each event prediction, the two loss functions are calculated as:
\begin{equation}
  \label{eqn:loc_loss}
\mathcal{L}_{\mathrm{loc}}=\sum_{i}^{N}\left(\lambda_{\mathrm{IOU}}
\mathcal{L}_{\mathrm{IOU}}\left(b_{i}, \hat{b}_{\hat{\sigma}(i)}\right)+\lambda_{\mathrm{L} 1}\left\|b_{i}-\hat{b}_{\hat{\sigma}(i)}\right\|_{1}\right)
\end{equation}
\begin{equation}
  \label{eqn:cls_loss}
\mathcal{L}_{\mathrm{cls}}=\frac{1}{N} \sum_{i=1}^{N}-\log \hat{p}_{\hat{\sigma}(i)}\left(c_{i}\right)
\end{equation}
where $\lambda_{\mathrm{IOU}}$ and $\lambda_{\mathrm{L1}}$ are weights for Intersection Over Union (IOU) loss \cite{yu2016unitbox} and L1 loss. \\
\textbf{Clip-level loss.} The audio tagging loss is defined as the binary cross-entropy between the clip-level class label $l_{\mathrm{tag}}$ and predicted audio tagging $y_{\mathrm{tag}}$:

\begin{equation}
  \label{eqn:cls_loss}
\mathcal{L}_{\mathrm{at}}=\mathrm{BCE}\left({l}_{\mathrm{tag}}, {y}_{\mathrm{tag}}\right)
\end{equation}
\vspace{-0.4cm}
\subsection{Self-supervised learning for SEDT}
\label{ssec:spsedt}
To better use the unlabeled or external datasets, such as AudioSet and SINS, we adopt a self-supervised learning method to pre-train SEDT on unlabeled data, which is named as Self-supervised Pre-training SEDT (SP-SEDT). Specifically, we randomly crop spectrogram along the time axis to obtain several patches, and then pre-train the model to predict corresponding locations of the patches. To preserve the category information in SP-SEDT, classification loss and feature reconstruction loss are also adopted as sub-objective terms. By means of such pre-text task, we hope that SEDT can localize sound event and maintain most category-related features at the same time. More details can be found in \cite{ye2021sp}.
\vspace{-0.2cm}
\subsection{Semi-supervised learning for SEDT}
\label{ssec:sssedt}
Pseudo-labelling \cite{lee2013pseudo} is one of the mainstream approaches of semi-supervised learning. It requires a well-trained model to generate pseudo labels on unlabeled data, so that in the next stage, the converged model can be re-optimized on both labelled data and unlabeled data jointly. Based on that, we propose an improved pseudo-labelling method for the Semi-Supervised learning of SEDT (SS-SEDT). SS-SEDT splits the training process into two stages: the burn-in stage and the teacher-guided stage. In the burn-in stage, SEDT is simply trained on the labeled dataset to initialize the model. At the beginning of the teacher-guided stage, the initialized model is copied into two models (a student model and a teacher model), and then the teacher model generates pseudo labels on unlabeled data so that the student model can gain knowledge from both labeled data and unlabeled data. To guarantee the quality of the pseudo labels, we revisit the following off-the-shelf techniques, and apply them in the teacher-guided process. The detailed training process of teacher-guided stage is shown in Algorithm~\ref{algorithm:tg}.
\begin{itemize}
\item {\bf EMA}: Unlike previous methods supervised by offline pseudo labels, we resort to a progressing teacher model to generate pseudo labels. The teacher model is updated from the student model through EMA and thus can be viewed as implicit ensemble models and provide more reliable guidance. Notice that although the usage of EMA is similar to that in the mean-teacher framework, the proposed method is different since pseudo labels involved are hard ones and no consistency loss is adopted.
\item {\bf Asymmetric augmentation}: Asymmetric augmentation has been introduced into semi-supervised image recognition \cite{sohn2020fixmatch} and SED \cite{Chan2021macaron}. Inspired by that, we adopt similar idea in the teacher-guided stage, during which weakly-augmented (frequency mask and frequency shift) spectrograms are fed into the teacher model to get pseudo labels and the student model make predictions on the strongly augmented (frequency mask, frequency shift, time mask and gaussian noise) version of the same data batch.
\item {\bf Mixup} \cite{zhang2017mixup}: We mix labeled data with ground-truth and unlabeled data with pseudo annotations together, which is supposed to improve the model robustness to pseudo annotation noise and alleviate the overfitting problem in model training. 
\item {\bf Focal loss} \cite{lin2017focal}: Focal loss is adopted to handle the unbalanced event categories in SED, without which the model may be overwhelmed by easily classified samples and produce biased outputs. It should be noted that focal loss is merely used in the teacher-guided stage, we believe such curriculum learning pattern may help our model learn from easy to difficult. 
\end{itemize}
\begin{algorithm}[t]
\caption{Pseudocode for teacher-guided stage}\label{algorithm:tg}
\SetKwInOut{Input}{Require}\SetKwInOut{Output}{Ensure}
\Input{$\mathcal{B}_L=\text{labeled batch, } \mathcal{B}_U=\text{unlabeled batch}$}
\Input{$S_{\theta}(x)=\text{student model}, T_{\theta^{\prime}}(x)=\text{teacher model}$}
\Input{$A_{w}(x)=\text{weak augmentation function}$}
\Input{$A_{s}(x)=\text{strong augmentation function}$}
\Input{$\alpha=\text{learning rate}, \gamma=\text{EMA ratio}$}
\Input{$\mathcal{L}=\text{loss function}$}
\Output{$\theta,\theta^{\prime}$}
\For{$i\rightarrow 1$ \KwTo $\text{max\_epochs}$}{
    \ForEach{$\mathcal{B}_L \cup \mathcal{B}_U \in \mathcal{B}$}{
        $\mathcal{J}_{\text {sup}} \leftarrow \frac{1}{\left|\mathcal{B}_{L}\right|} \sum_{(x_{i},y_{i}) \in \mathcal{B}_{L}} \mathcal{L}\left(S_{\theta}\left(A_{w}(x_{i})\right), y_{i}\right)$\;
        \lForEach{$x_i \in \mathcal{B}_U$}{$y_i \leftarrow T_{\theta^{\prime}}(A_{w}(x_i))$}
        $\hat{\mathcal{B}}\leftarrow \text Mixup(\mathcal{B}_L, \mathcal{B}_U)$\;
        $\mathcal{J}_{\text {unsup}} \leftarrow \frac{1}{\left|\hat{\mathcal{B}}\right|} \sum_{(\hat x_{i},\hat y_{i}) \in \hat{\mathcal{B}}} \mathcal{L}\left(S_{\theta}\left(A_{s}(\hat x_{i})\right), \hat y_{i}\right)$\;
        $\theta \leftarrow \theta - \alpha (\frac{\partial{\mathcal{J}_{\text {sup}}}}{\partial \theta} + \frac{\partial{\mathcal{J}_{\text {unsup}}}}{\partial \theta})$\; 
        $\theta^{\prime} \leftarrow \gamma \theta^{\prime} + (1 - \gamma) \theta$\;
        }
}
\end{algorithm}
\section{Frame-wise CNN model}
\label{sec:mlflcnn}
The pipeline of the frame-wise CNN model is illustrated in Fig.~\ref{fig:mlflcnn_overview}. At first, MLFL-CNN is preliminarily trained with weakly labeled data and strongly labeled synthetic data to acquire basic event detection and audio tagging ability. Then, it attaches pseudo strong labels to the weakly-labeled and unlabeled data, and the model is jointly trained with all these data in a self-training manner. Finally, the trained MLFL-CNN provides audio tags and strong pseudo labels for the weakly-labeled data and unlabeled data to train the tag-conditioned CNN \cite{Ebbers2020}, which gives the final SED results.
\begin{figure}[h]
  \centering
  \centerline{\includegraphics[width=\columnwidth]{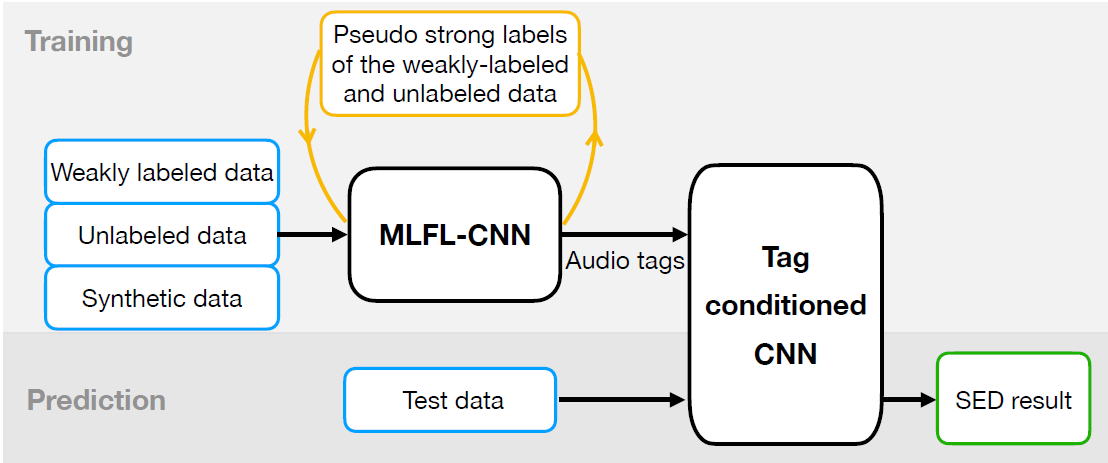}}
  \caption{Overview of the frame-wise model}
  \setlength{\belowcaptionskip}{-0.5cm}
  \label{fig:mlflcnn_overview}
\end{figure}

The MLFL-CNN model contains three branches. The first branch is the embedding-level attention pooling branch based on the MIL framework, which is the same with \cite{lin2020specialized}. The second branch is the sound event detection branch which is introduced to exploit the strong labels of synthetic data and uses focal loss as its supervision. The third branch is the domain adaptation branch which uses metric learning by inter-frame distance contrastive loss, more details of which can be found in \cite{huang2020learning}. During training process, the MLFL-CNN adopts the mean-teacher architecture and pseudo-labelling framework simultaneously \cite{rui2022couple}. It combines clip-level loss (for weakly-labeled data), frame-level loss (for data with strong labels and pseudo strong labels), inter-frame distance contrastive loss (for real data and synthetic data), and consistency loss together. And the tag-conditioned CNN takes spectrograms and audio tags predicted by the MLFL-CNN as inputs, and uses the strong labels of synthetic data and pseudo strong labels of real data as ground-truth to train.
\vspace{-0.2cm}
\section{Fusion of the two models}
\label{sec:fusion}
\subsection{Preliminary: Class-specific PSDS}
\label{ssec:class_psds}
The essence of PSDS is to obtain a function $r(e)$ of effective TP rate (\text{eTPR}) changing with effective FP rate (\text{eFPR}), and calculate the integral of this function over $(0, e_{\max})$, where $e_{\max}$ represents the maximum value of \text{eFPR} value \cite{bilen2020framework}. We notice that the original calculation of \text{eTPR} relies on two class-averaged indicators $\mu_{\mathrm{TP}}$ and $\sigma_{\mathrm{TP}}$. To decouple the eTPR according to the event category, we simply replace class-averaged indicators with class-dependent ones and finally redefine the PSDS value of given category as follow:
\begin{equation}
  \label{eqn:tp_c}
\mu_{\mathrm{TP, c}}=r_{\mathrm{TP}, c}  \quad \sigma_{\mathrm{TP, c}}= r_{\mathrm{TP}, c}-\mu_{\mathrm{TP, c}}
\end{equation}
\begin{equation}
  \label{eqn:etpr_c}
\text {eTPR}_c: \quad r_c(e) \triangleq \mu_{\mathrm{TP, c}}(e)-\alpha_{S T} * \sigma_{\mathrm{TP, c}}(e)
\end{equation}
\begin{equation}
  \label{eqn:psds_def}
\mathrm{PSDS}_c \triangleq \frac{1}{e_{\max }} \int_{0}^{e_{\max }} r_c(e) d e
\end{equation}
where $\mathrm{PSDS}_c$, $\text {eTPR}_c$, $\mu_{\mathrm{TP, c}}$ and $\sigma_{\mathrm{TP, c}}$ are corresponding class-wise indicators for specific event class $c$.
\subsection{Model fusion method}
\label{ssec:model_fusion}
The core of model fusion is to calculate the class-wise fusion coefficients of each model's prediction during the evaluation stage. Assume that there are $N$ models $m_i(i = 1,2,\dots N)$, for each sound event class $c$, the PSDS of model $m_i$ on $c$ is denoted as $\mathrm{PSDS}_{i,c}$. Then the fusion coefficient of model $i$ on category $c$ is defined as:

\begin{equation}
  \label{eqn:w_ensemble}
w_{i,c} = \frac{\mathrm{PSDS}_{i,c}}{\sum_{i = 1}^{N}\mathrm{PSDS}_{i,c}}
\end{equation}

Therefore, for specific event category $c$, the final fusion probability $\hat{p_c}$ is formulated as the weighted linear combination of each model’s predicted probability  $p_{i,c}$:

\begin{equation}
  \label{eqn:p_ensemble}
\hat{p_c} = \sum_{i = 1}^{N}w_{i,c}* p_{i,c}
\end{equation}

It is noteworthy that the above PSDS in Eq.(7) can be interpreted as PSDS1 or PSDS2 for this year's DCASE task4, so two different sets of parameters $w_{i,c}$ can be obtained on the development set and utilized to improve PSDS1 and PSDS2 respectively. 
\section{Post-Processing}
\vspace{-0.2cm}
\label{sec:post-processing}
In order to reduce the noise in frame-level probability and make sound events continuous, it is necessary to perform a smoothing operation, such as mean filter or median filter, on the frame-level probability. Currently, median filtering with a fixed window length or with the average length of each event calculated on the development set is generally utilized \cite{lin2019guided}. In this paper, we perform median filtering and mean filtering (with larger window size) on frame-level probabilities in sequence, and propose a method to search for optimal class-wise window lengths on the development set. 
\begin{table}[htb]
	\centering
	\setlength{\abovecaptionskip}{0.cm}
	\setlength{\belowcaptionskip}{0.05cm}
	\caption{The PSDS on the validation set}
	\label{table:psds_system}
	\setlength{\tabcolsep}{4.5mm}{
	\begin{tabular}{cccc}
		\toprule  
		System&Extra data&PSDS1&PSDS2 \\ 
		\midrule  
		Baseline 1&&0.336&0.536 \\
		Baseline 2&\Checkmark&0.351&0.552 \\
		System 1&\Checkmark&0.449&0.645 \\
		System 2&\Checkmark&0.115&0.816 \\
		System 3&&0.420&0.618 \\
		System 4&&0.099&0.783 \\
		\bottomrule  
	\end{tabular}}
\end{table}
\vspace{-0.8cm}
\begin{table}[htb]
	\centering
	\setlength{\abovecaptionskip}{0.cm}
	\setlength{\belowcaptionskip}{0.05cm}
	\caption{Ablation study on techniques in SS-SEDT}
	\label{table:ablation_sssedt}
	\setlength{\tabcolsep}{3mm}{
	\begin{tabular}{cccccc}
		\toprule  
		MU&FL&AA&EMA&PSDS1&PSDS2 \\ 
		\midrule  
		&\Checkmark&\Checkmark&\Checkmark&0.372&0.570 \\
		\Checkmark&&\Checkmark&\Checkmark&0.349&0.540 \\
		\Checkmark&\Checkmark&&\Checkmark&0.369&0.566 \\
		\Checkmark&\Checkmark&\Checkmark&&0.357&0.538 \\
		\Checkmark&\Checkmark&\Checkmark&\Checkmark&0.388&0.573 \\
		\bottomrule  
	\end{tabular}}
\end{table} 

Specifically, for a given event class $c$, we enumerate window length $wl_c$ from 1 to 500, and find the optimal length ${wl_c}^*$ to optimize PSDS1 and PSDS2 respectively:
\begin{equation}
  \label{eqn:wl_c}
{wl_c}^* = \mathop{{\arg\max}}_{wl_c}  \frac{\mathrm{PSDS}_c}{\mathrm{PSDS}}
\end{equation}

Finally, in the hybrid system, the event-level predictions of SEDT are firstly obtained in an end-to-end manner and then converted into frame-level probabilities, before being fused with frame-wise model and finally post-processed to get the ultimate results. 
\vspace{-0.2cm}
\section{Experiment}
\vspace{-0.15cm}
\label{sec:exp}
\subsection{Experiment Setup}
\label{ssec:exp_setup}
For SEDT not using external data, we firstly pre-train it on unlabeled real subset (14412 clips), then simply train it on the weakly labeled training set (1578 clips) and synthetic 2019 subset (2045 clips) during burn-in stage, and finally use weakly labeled set, synthetic 2019 subset, synthetic 2021 subset (10000 clips), and unlabeled subset to conduct teacher-guided learning. For SEDT using external data, the two main differences compared to the above lie in 1) models are pre-trained on both unlabeled real subset and SINS subset (72894 clips), 2) an additional strongly labeled set (3470 clips) is further included in the teacher-guided stage. The detailed settings can be found in our repository \footnote{\url{https://github.com/Anaesthesiaye/sound_event_detection_transformer}}.

For frame-wise model not using external data, the training set contains the weakly labeled training set, the unlabeled training set, and synthetic 2021 subset. While for systems using external data, we add the same strongly labeled set taken from AudioSet to the original strong labeled set. The detailed settings of training hyper-parameters  and configurations can be found in \cite{Tian2021}.
\vspace{-0.1cm}
\subsection{Results of Submitted Systems}
\vspace{-0.05cm}
\label{ssec:comp_res}
Table~\ref{table:psds_system} shows the performance of our submitted systems, all of which are fused models of ensemble frame-wise CNN models and ensemble SEDT. Among them, system 1 and 2 incorporate external data, while system 3 and 4 do not. Besides, model fusion and window tuning methods proposed in Section 4 and Section 5 are utilized in system 1, 3 to improve their PSDS1 and in system 2, 4 to improve their PSDS2 separately. As shown in Table~\ref{table:psds_system}, our hybrid systems outperform the official baseline considerably whatever the usage of external data. Moreover, our systems ranked 6th / 9th in the challenge respectively. While they are inferior to the winner models, our designed components are orthogonal to network architecture and data augmentation, which means that they may generalize to other models and bring about promising improvements.
\begin{table}[htb]
	\centering
	\setlength{\abovecaptionskip}{0.cm}
	\setlength{\belowcaptionskip}{0.05cm}
	\caption{Ablation study on window tuning and model fusion}
	\label{table:ablation_tuning_fusion}
	\setlength{\tabcolsep}{2.7mm}{
	\begin{tabular}{cccccc}
		\toprule  
		Id&Model&MF&WT&PSDS1&PSDS2 \\ 
		\midrule  
		1&Single SEDT&&&0.415&0.582 \\
		2&Ensemble SEDT&&&0.431&0.607 \\
		3&Single frame&&&0.349&0.668 \\
		4&Ensemble frame&&&0.392&0.673 \\
		5&Hybrid system &\Checkmark&&0.437&0.740 \\
		6&Hybrid system &\Checkmark&\Checkmark&0.449&0.816 \\
		\bottomrule  
	\end{tabular}}
\end{table} 
\vspace{-0.7cm}
\subsection{Ablation Study}
\label{ssec:abla_study}

\textbf{Techniques in SS-SEDT. } To verify the effectiveness of techniques in SS-SEDT, we conduct ablation study using single SS-SEDT model without external data. Table~\ref{table:ablation_sssedt} shows the results of models trained without specific technique, where MU, FL, AA denotes Mixup, Focal Loss, Asymmetric Augmentation mentioned in Section 2.4 respectively, and the model trained without AA means that the inputs of teacher and student model are both weakly augmented. It can be seen that all techniques can improve the performance of SS-SEDT and it can finally reach a PSDS1 of 0.388 and a PSDS2 of 0.573 while incorporating all techniques.\\
\textbf{Window tuning and model fusion.} To investigate the effects of window tuning and model fusion strategy, we conduct ablation study using SEDT and frame-wise model trained with external data. Table~\ref{table:ablation_tuning_fusion} compares the performance between models under different settings. In the above table, MF and WT denote Model Fusion and Window Tuning methods proposed in Section 4 and 5 respectively, and frame-wise model is abbreviated to ``frame''. Among all these models, model 2 and 4 are ensemble models of top 1-5 single models, while hybrid system represents the fused model of ensemble SEDT and ensemble frame-wise model. By comparing model 1, 2 with model 3, 4, it is obvious that SEDT can achieve higher PSDS1 while frame-wise model is better at PSDS2. Moreover, by comparing model 5 with model 2, 4, we can see that while SEDT and frame-wise model have their own edges, they can complement each other, since the hybrid system achieve further improvements compared to single ensemble models. By comparing model 6 with model 5, the effectiveness of window tuning can be validated, since model 6 provides the best PSDS1 (0.449) and PSDS2 (0.816). 
\vspace{-0.2cm}
\section{Conclusions}
\vspace{-0.2cm}
\label{sec:conclusion}

In this paper, we developed a framework to fuse the detection results of the frame-wise model and event-wise model, which leads to an improved PSDS1 of 0.420 and PSDS2 of 0.783 on the validation set compared to individual ensemble models. 

\vspace{-0.2cm}
\section{ACKNOWLEDGMENT}
\vspace{-0.2cm}
\label{sec:ack}

This work was partly supported by the National Natural Science Foundation of China (62276250) and the Beijing Natural Science Foundation (Z190020).

\bibliographystyle{IEEEtran}
\bibliography{refs}

%
%
%
%
%
%
%
%
%

\end{sloppy}
\end{document}